\documentstyle[twoside,fleqn,ams,espcrc2,graphicx,epsfig]{article}
\newcommand{\AmS}{{\protect\the\textfont2  
  A\kern-.1667em\lower.5ex\hbox{M}\kern-.125emS}}
\newcommand{\ct}{\cite}

\title{Multiplicities, fluctuations and correlations\thanks{
      Talk given at the 31st International Conference on High Energy
      Physics
      (ICHEP 2002), Amsterdam, The Netherlands, 24 -- 31 July 2002}}
\author{Edward K.G. Sarkisyan\address{EP Division, CERN, CH-1211
        Geneva 23, Switzerland}\address{Physics Department, 
University of Antwerpen, Universiteitsplein 1, B-2610 Wilrijk, Belgium
}%
 \thanks{Electronic address: {\tt sedward@mail.cern.ch}}}
\begin{document}

\begin{abstract}
The recent results on hadron multiplicities in heavy and light quark
fragmentation, 
multiplicity 
local fluctuations and multiparticle correlations submitted to the
Conference are
reviewed.
\vspace{1pc}
\end{abstract}
\maketitle

\section{Introduction}

One of the most important observables in particle production processes is
{\it multiplicity}, i.e.  the number of particles (mostly, hadrons)
produced in the collision \cite{hmrep}. 
The multiplicity dependence on  the energy scale, species of particles,
event flavour content  are among the main predictions of the
theory of strong
interactions,  quantum chromodynamics (QCD) \cite{hmrep,qcdrev}. On the
other
hand, the
multiplicity is used  to select or to describe events, e.g.
as a trigger for specific
processes, as an input for kinematic
variables' spectra.
The distribution of
multiplicity, its mean value and multiplicity fluctuations are  the
essential
characteristics of the 
collision dynamics. 
However, the multiplicity distribution tells us
just about the
averaged, integrated numbers, while deeper information comes
from the 
moments of the  distribution, which measure  
particle {\it correlations}, i.e. probe the interaction dynamics 
\cite{cfrep}.  

Here, I report on the multiplicity flavour dependence\footnote{Other 
aspects of multiplicity studies such as
multiplicity energy
dependence,
3-jet
multiplicity, multiplicities of quark and gluon jets etc. are reviewed by
M.
Siebel \ct{fragm}.},
 on the analyses of local multiplicity fluctuations
and multiparticle correlations\footnote{
Bose-Einstein correlation studies   are
reviewed by \v{S}. Todorova-Nov\'a \ct{bec}.}. 
These studies provide us
with details of
strong
interactions and  allow us to estimate 
the level of the applicability of QCD, based on the  partonic picture, to 
the production of hadrons which are the  
experimentally observed objects.

\section{Definitions and notations \cite{hmrep,cfrep}}

The multiplicity distribution, or the density $\rho_n$, of multiplicity
$n$
of particles
with kinematic  variables $p_1,p_2,\ldots, p_n$ is defined by 
the inclusive probability spectrum,
\vspace*{-.2cm}
$$
\rho_n(p_1,p_2,\ldots, p_n)=\frac{1}{N_{\rm ev}}\,
\frac{dn(p_1,p_2,\ldots, p_n)}{dp_1 dp_2\cdots dp_n}\:, 
$$
where $N_{\rm ev}$ is the number of events. 

As it follows from this formula, the single particle distribution
$\rho_1(p_1)$ gives an {\it average} multiplicity,
$
\langle n\rangle=\int \rho_1(p_1) dp_1
$,
while integration of the $q$-particle densities leads to the unnormalised
$q$th order {\it
factorial
moments},
\vspace*{-.2cm}
$$
\hspace*{-.84cm}
f_q=\int
\rho_1(p_1,p_2\ldots p_n) dp_1dp_2\cdots dp_n
\vspace*{-.075cm}
$$
\begin{equation}
\label{ufm}
\end{equation}
\vspace*{-.65cm}
$$
=\langle n(n-1)\cdots
(n-q+1)\rangle \equiv  
\langle n^{[q]}\rangle\:.
$$

\noindent 
The  {\it normalised} factorial  moments are then defined as  
$F_q=f_q/\langle n\rangle ^q$.

The $q$-particle densities give us a way to study particle 
correlations  described by the  $q$-particle correlation functions, 
(factorial) {\it cumulants}, 
$C_q(p_1,\ldots , p_q)$. The cumulants  vanish whenever one of their
arguments is
statistically independent, i.e. these functions measure
{\it genuine} $q$-particle
correlations. 

The cumulants are constructed from the multiplicity densities, e.g. 
\vspace*{-.2cm}

$$
C_1(p_1)=\rho_1(p_1),
$$
$$
C_2(p_1,p_2)=\rho_2(p_1,p_2)-\rho_1(p_1)\rho_1(p_2),
$$
$$
C_3(p_1,p_2,p_3)=\rho_3(p_1,p_2,p_3) -$$
\begin{equation}
\label{ucm}
\hspace*{.15cm}
\sum_{(3)}\rho_1(p_1)\rho_2(p_2,p_3)+
2\rho_1(p_1)\rho_1(p_2)\rho_1(p_3)\:.
\end{equation}
These functions, being properly normalised, are used to study
{\it multi}particle correlations in different kinematic variables.

In studies of local fluctuations and
correlations, i.e those in phase-space bins, one uses the
{\it bin-averaged} factorial moments and cumulants.
The
phase space is divided into $M$ bins of equal size, so that  $\langle
n_m \rangle$ is the number of particles in the $m$th bin.  
Then, the normalised bin-averaged factorial moment is defined via 
Eq. (\ref{ufm}) averaged over bins, 

\begin{equation}
\label{nfm}
\hspace*{.2cm}
F_q(M)={1\over M}\sum_{m=1}^M\frac{f_q(m)}{\langle n_m \rangle ^q}\equiv 
{1\over M} \sum_{m=1}^M
\frac{\langle n_m^{[q]}\rangle}{\langle n_m \rangle ^q}. 
\end{equation}

\noindent
These factorial moments
 have been
extensively used to  extract the {\it non-statistical} (non-Poissonian)
fluctuations in many types of collisions \cite{cfrep}. Such fluctuations
lead to the power-law scaling  of factorial moments as a function
of $M$ called the {\it intermittency} phenomenon. 
 
In analogy with Eq. (\ref{nfm}), Eqs. (\ref{ucm}) are averaged
to the 
bin-averaged normalised cumulants,
\begin{equation}
\label{ncm}
\hspace*{1.7cm}
K_q(M)=
{1\over M} \sum_{m=1}^M
\frac{\langle k_q^{(m)}\rangle}{\langle n_m \rangle ^q},
\end{equation}
\vspace*{-.15cm}
where $ k_q^{(m)}$ are the Mueller moments, the functions of
unnormalised moments $f_q(m)$, Eq. (\ref{ufm}), 
\vspace*{-.15cm}
$$
k_1=\langle n_m\rangle, \quad k_2^{(m)}=\langle n_m^{[2]}\rangle -\langle
n_m\rangle ^2,
$$
\begin{equation}
\label{mum}
\hspace*{.75cm}
k_3^{(m)}=\langle n_m^{[3]}\rangle - 3\,\langle n_m^{[2]}\rangle \langle
n_m\rangle\,
+ 2\, \langle n_m\rangle ^3.
\end{equation}

\section{Hadronisation of heavy and light quarks}

In this Section, I consider the recent results on
hadron multiplicities from  fragmentation of heavy
and
light quarks 
\ct{delbl,opallq,opalbl}.
The study of the quark content in multiparticle
production provides 
one of the basic tests of QCD. 
The results from LEP are of a special interest since they 
cover a wide centre-of-mass energy region and can be directly compared
with
QCD which is mostly predictable at the asymptotic energies \ct{qcdrev}.      

\begin{figure}[t]
\vspace{7pt}
\includegraphics[width=73mm]{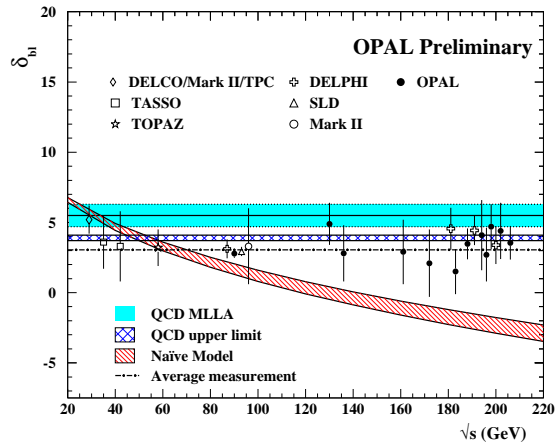}
\vspace*{-0.8cm}
\caption{
The difference in the mean charged multiplicities,
$\delta_{\rm bl}$, 
between heavy and light quark pairs as a function of centre-of-mass
energy. The dashed-dotted line is the combined result from all
measurements, see Ref. \ct{opalbl} for more 
details.}
\label{fig:opalbl} 
\vspace*{-.8cm}
\end{figure}

In Refs. \ct{delbl} and \ct{opalbl}, a study of the fragmentation of
heavy b-quark and light quarks (l = u, d, s) is performed. The
measurements of
the difference in charge particle multiplicities,
$\delta_{\rm bl} = \langle n_{{\rm b}{\bar {\rm b}}}\rangle -
                   \langle n_{{\rm l}{\bar {\rm l}}}\rangle$,
 for ${\rm b}{\bar {\rm
b}}$ and ${\rm l}{\bar {\rm
l}}$ events in e$^+$e$^-$ annihilation at the
centre-of-mass energies above the Z$^0$ peak
are carried out. The findings are compared to the theoretical predictions
of
QCD and to a more phenomenological (the so-called na\"\i ve) model (for
review,
see Ref. \ct{qcdrev}). The QCD calculations predict energy-independent   
behaviour of the multiplicity difference $\delta_{\rm bl}$, while in the
na\"\i ve model one expects the decrease with increasing energy. The
latter
is connected with the assumption that the hadron multiplicity
accompanying the heavy hadrons in ${\rm b}{\bar {\rm b}}$ events is the
same as the multiplicity in ${\rm l}{\bar {\rm l}}$ events at the energy
left to the system once the heavy quarks have fragmented.

The difference between the heavy and light 
mean quark-pair multiplicities obtained by DELPHI
at 206
GeV is $\delta_{\rm bl}=4.50\pm1.05({\rm stat})\pm0.52({\rm syst})$
\ct{opalbl}, while OPAL finds 
 $\delta_{\rm bl}=3.44\pm0.40({\rm stat})\pm0.79({\rm syst})$ in the
energy
range of 130 -- 206
GeV \ct{opalbl}. The difference in the values is connected with some
differences in the data processing procedure. In the meantime, the results
of the energy-dependence is found to be the same for the both studies: the
mean
multiplicity difference is energy-independent and favours the QCD
predictions while it is inconsistent with the flavour-independent na\"\i
ve
model (Fig. \ref{fig:opalbl}). Also in agreement with QCD 
is the ratio between light quark multiplicities, $\langle n_{\rm
i}\rangle:\langle n_{\rm
j}\rangle \sim 1$, where i,j =\{u,d,s\}, as obtained by OPAL
at the Z$^0$ peak energy \ct{opallq}. To note is that the $\langle n_{\rm
u}\rangle$ and $\langle n_{\rm
d}\rangle$ are  highly statistically anti-correlated ($-90$\%) which is
due
to their fractions in K-mesons.

\section{Scaling of local fluctuations in  hadronic Z$^0$ decays}

During the last decades, the phase-space local multiplicity fluctuations 
are
actively studied in many reactions, from leptonic to nuclear
collisions, and the intermittency scaling of the fluctuations has been
established \ct{cfrep}. 

All the studies show that the intermittency is more pronounced  
in high dimensions, while in one dimension (e.g., in rapidity)  
the effect
is diluted by projection onto one dimension. This leads to flattenning of
the factorial moments.  Such a behaviour is well understood
from a
QCD parton shower which is a three-dimensional branching and naturally
leads
to the fractality. The best area to study the effect and its connection
with QCD is given by e$^+$e$^-$ annihilation, where such investigations
has
been  performed earlier \ct{cfrep,del3d,opal3d} and new analysis is
carried out by
L3 \ct{l3sa}.
 
The L3 analysis gives us further hints about the intermittency origin,
which is currently far from understanding. The new study employs  the
fact that so far the fluctuations in many dimensions were studied with the
same number $M$ of bins  in each dimension/direction. This leads to 
 self-{\it similar} fractals, i.e. the fluctuations in any direction
are assumed to be the same, or {\it isotropic}. In case when the dynamics
in different
directions are not equivalent, the fluctuations become  {\it anisotropic}
and  this tells us about  the self-{\it affinity} \ct{liu}.
A self-affine behaviour has been observed in hadronic interactions
\ct{liu}, while
a self-similar scaling has been qualitatively confirmed in e$^+$e$^-$
annihilation \ct{opal3d}. In terms of factorial moments, Eq.(\ref{nfm}),
the 3-dimensional moments are expected to exhibit the intermittency
property
in e$^+$e$^-$ collisions when phase-space is partitioned isotropically,
while this scaling in hadronic interactions only occurs for anisotropic
partitioning.

\begin{figure}[t]
\vspace{-7pt}
\includegraphics[height=82mm,width=82mm]{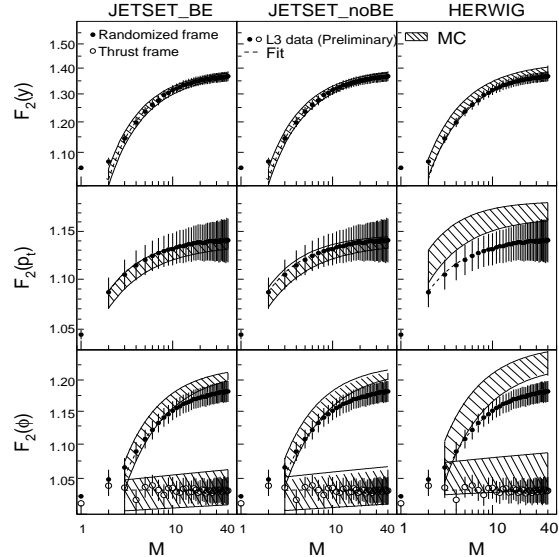}
\vspace*{-1.5cm}
\caption{
The 1-dimensional 2nd-order factorial moments, Eq. (\ref{nfm}), as a
function of the number $M$ of bins in the randomised and the 
thrust frames compared to the Monte Carlo predictions with and without
Bose-Einstein correlations. The fit is given by the first f-la of Eqs.
(\ref{1df2}). The behaviour in the q${\bar {\rm q}}$ frame is similar
to that in the randomised frame. See
text and \ct{l3sa} for details.}
\vspace*{-.6cm}
\label{fig:l3sa} 
\end{figure}

In order to quantitatively study the observed isotropic fluctuations,
the method of the Hurst exponent is used \ct{liu}.   
The Hurst exponent, 
\begin{equation}
\label{hurst}
\hspace*{2.cm}
H_{ab}={\ln M_a}/\,{\ln M_b},
\end{equation}
is obtained from the fit of the 
one-dimensional 
second-order
factorial moments,
\begin{equation}
\label{1df2}
F_2(M_a)=A_a+B_a\,M_a^{-\gamma_a}\: \Rightarrow\: H_{ab}=
\frac{1+\gamma_b}{1+\gamma_a}.
\end{equation}
Here $a$, $b$ are the directions of the $(a,b)$ plane.
For the isotropic dynamical fluctuations, 
$H_{ab}= 1$, while $H_{ab}\neq 1$ if the fluctuations are anisotropic.

L3 uses rapidity $y$, azimuthal angle $\varphi$ and transverse
momentum $p_t$ -- the
variables often used in multiparticle studies -- in the analysis. The
variables are  defined with respect to the thrust axis, as well as the 
other two frames are considered: the randomised and the q${\bar {\rm q}}$
frames.
In the former one, the $\varphi$-angles are randomly chosen
being uniformly distributed in $[0,2\pi]$, while in the latter case one 
uses
Monte Carlo to
correct the original thrust axis to the q${\bar {\rm q}}$ axis.

\begin{figure}[t]
\includegraphics[height=88.3mm,width=82mm]{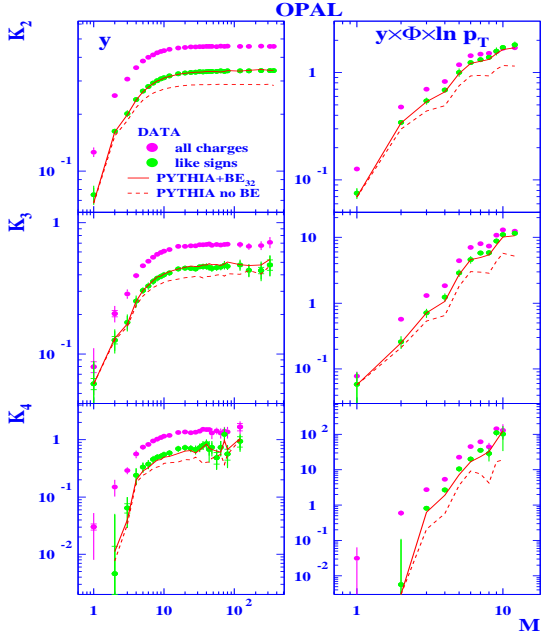}
\vspace*{-1.55cm}
\caption{
The all-charge and like-sign cumulants, Eq. (\ref{ncm}), as a
function of the number $M$ of bins in rapidity and in three
dimensions. The like-sign cumulants are   
compared to the Monte Carlo predictions with and without
Bose-Einstein correlations. 
See text and Ref. \ct{opalgc} for more details.}
\vspace*{-.9cm}
\label{fig:opal13c} 
\end{figure}

The 3-dimensional factorial moments are found 
to exhibit the intermittency scaling
when  phase-space is
isotropically partitioned. This confirms earlier 
indications for isotropic fluctuations in  e$^+$e$^-$
annihilations
\ct{del3d,opal3d}.  No sensitive differences in the different frames
as well as no
disagreements between the data and Monte Carlos are observed. 

Fig. \ref{fig:l3sa} shows the 2nd-order factorial moments in one
dimension,
from which the quantitative test of the isotropy of the fluctuations is
made
using Eqs. (\ref{1df2}). The calculations of the Hurst exponent give:
$H_{ab}\approx 1$, $a,b=\{y,\varphi,p_t\}$, i.e. due to the 
definition of Eq. (\ref{hurst}) 
 one concludes about the equivalence of the directions
and about the isotropy of the fluctuations.   The fluctuations in
$\varphi$ are nearly absent in the
thrust frame. In all frames the data are well described by Jetset, while
somewhat less well by Herwig. No sensitivity to Bose-Einstein correlations
is seen. 
 From its study, L3 also concludes  that there is the dependence on the
QCD
dynamics which serves to decrease the fluctuations in the thrust frame.

The 2-jet sub-samples are also analysed using the
Durham jet algorithm. The fluctuations in
the 2-jet events show  the self-affine (anisotropic)  fluctuations. This
looks like the hard gluon emission leads to the isotropy.

\section{Multiparticle correlations}

The multiparticle correlation studies
  are performed for e$^+$e$^-$
annihilations at the Z$^0$ energy \ct{opalgc} and for 
${\bar {\rm p}}$p
annihilation
 at the incident momentum of  22.4 GeV/$c$ \ct{bpp}.  
      
Genuine multiparticle correlations in e$^+$e$^-$ annihilations are studied
by OPAL
\ct{opalgc}. To extract the correlations, the method of the
bin-averaged normalised cumulants is used, see Eqs.
(\ref{ncm})--(\ref{mum}). In addition to the earlier analysis \ct{opal3d}
of the cumulants of 
all charged particles (``all-charge cumulants''), OPAL  studies the
cumulants of same-charge particles (``like-sign cumulants'').
The investigation is performed in rapidity $y$, 
azimuthal angle $\varphi$ and transverse
momentum $p_t$, the same variables as  in the above described L3 analysis,
but
the sphericity axis is used as a reference axis.

\begin{figure}[t]
\vspace{8pt}
\includegraphics[height=82mm,width=72mm]{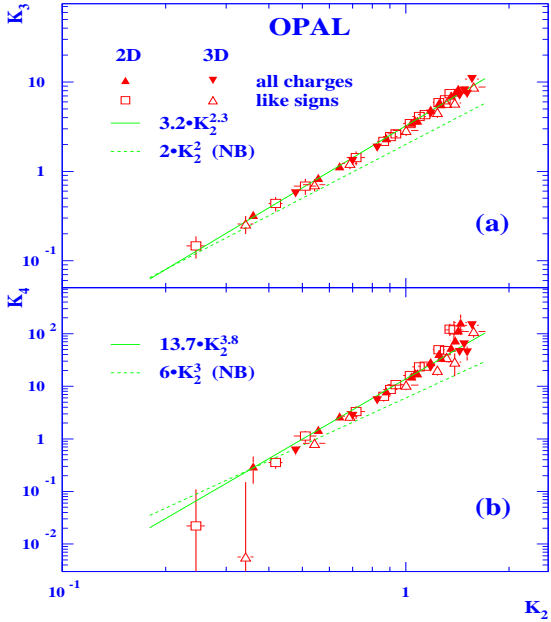}
\vspace*{-.8cm}
\caption{The Ochs-Wosiek plot in two and three dimensions for all-charge
and like-sign cumulants. The dashed line shows the Negative Binomial
prediction. The solid line is a fit, $\ln K_q=a_q+r_q\ln K_2$.  See text
and \ct{opalgc} for more details.}
\vspace*{-.7cm}
\label{fig:owk} 
\end{figure}

Fig. \ref{fig:opal13c} shows the all-charge and like-sign cumulants
calculated
in one (rapidity) and three dimensions. Even in one dimension, 
positive
genuine correlations of groups of two, three, four particles are
present. The cumulants exhibit a scaling behaviour, although in
rapidity the saturation already appears at the moderate bin sizes.
The like-sign
cumulants increase faster and 
 drive the all-charge ones at small bins while the 
unlike-sign cumulants are almost constant. This points to the
likely
influence of Bose-Einstein correlations.

The comparison between  the data and Monte Carlo like-sign cumulants
shows
that the model describes well the data when
Bose-Einstein correlations are implemented. The same is true for
the all-charge cumulants (not shown here).  This is not the case of the
L3 results 
where no difference is found for
the models with and without Bose-Einstein correlations, see  Fig.
\ref{fig:l3sa}. To note is that the OPAL data is described well in one
dimension as well as in three dimensions.

From Fig. \ref{fig:opal13c}, one can see that Bose-Einstein correlations
which are implemented  in the model  as the correlations of {\it two}
identical
particles and are pair-wise adjusted, well also describe the
cumulants of
$q>2$. This suggests to consider the interdependence of the 2nd- and
higher-order cumulants, Fig. \ref{fig:owk}.
The fit akin to the Ochs-Wosiek one \ct{cfrep} gives the same
parameter values for the all-charge and like-sign cumulants, while
disagrees
with the Negative Binomial predictions.

The multiparticle fluctuations are also studied in \ct{bpp} using the
Serpukhov fixed target   ${\bar {\rm p}}$p
annihilation data at  22.4 GeV/$c$. The differential spectra of $k$
particles in
pseudorapidity bins are analysed relative to the similar background
spectra. The dips in the distributions of such ratios are found to be
independent on the number $k$ of particles, $\sim 1.8$. The number of
clusters is estimated to be about 2-3 with two particles per cluster in
average.   
It is found that the data in the non-annihilation channel is similar to
that from the inelastic pp
collisions at 69 GeV/$c$ from the same accelerator.  These observations
are treated in  frames of different mechanisms, e.g. a model with mesons
emitted from intermediate nucleon isobars is suggested. 
\medskip

\noindent
I am thankful to the ICHEP02 Organising Committee, to convenors of the QCD
Session,
and to my colleagues at CERN and particularly  in OPAL  for giving me
the opportunity to make this presentation and 
for their kind help and support.

\end{document}